\documentclass[cameraready]{Interspeech}

\title{From A to B to A: Palindromic Zero-Shot Voice Conversion \\with Non-Parallel Data}

\author[affiliation={1}]{Moshe}{Mandel}
\author[affiliation={2}]{Shlomo E.}{Chazan}

\address{
    $^1$ Independent, Israel
    $^2$ OriginAI, Israel
}

\email{moshe.mandel@mail.huji.ac.il, shlomi@originai.co}

\keywords{voice conversion, synthetic parallel data}

\usepackage{xcolor}

\usepackage{comment}
\usepackage{cleveref}
\usepackage{multirow}
\usepackage{makecell}
\usepackage{graphicx}
\usepackage{booktabs}

\usepackage{placeins}
\usepackage{dblfloatfix}

\usepackage{subcaption}
\usepackage{pgfplots}
\pgfplotsset{compat=1.18} 
\usepgfplotslibrary{groupplots}

\pgfplotsset{
  legend image code/.code={
    \path[fill=\pgfplotslegendimagefill, draw=none]
      (0cm,-0.12cm) rectangle (0.35cm,0.12cm);
  }
}

\begin{document}

\maketitle

\begin{abstract}
    We present a voice conversion (VC) framework that utilizes K-Nearest Neighbors (KNN) retrieval over WavLM representations to align non-parallel source and target speech, constructing synthetic training pairs for supervised learning.
    The retrieved segments serve as synthetic inputs, while real target audio provides ground-truth outputs, forming a synthetic-to-real training paradigm that naturally supports multilingual data without requiring parallel corpora or explicit alignment.
    To ensure consistent target-speaker identity, we incorporate a speaker loss derived from a pretrained speaker verification model.
    Experiments across multiple languages demonstrate that the proposed approach achieves high naturalness and strong speaker similarity, outperforming competitive VC baselines, despite being trained exclusively on English data. Samples can be accessed at: \url{https://palindromic-vc.github.io}.

    
    
\end{abstract}

\section{Introduction}
\begin{figure}[t]
  \centering
  \includegraphics[width=\columnwidth]{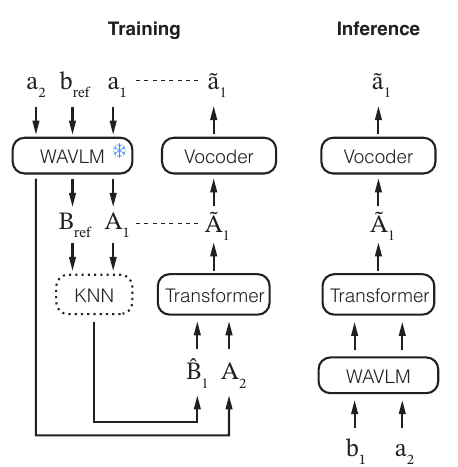}
  \caption{Overview of our palindromic voice conversion scheme. During training we synthesize input to the model by utilizing offline voice conversion via KNN of WavLM features. We optimize the latent and waveform outputs $\widetilde{A}_1$ and $\widetilde{a}_1$ against their supervised real counterparts, using a collection of reconstruction, adversarial and speaker losses. We generalize to real input at inference time. See project page for interactive figure.}
  \label{fig:overview}
\end{figure}

Voice conversion (VC) aims to modify the speech of a source speaker to match the characteristics of a target speaker while preserving the linguistic content. These characteristics may include speaker identity, prosody, or emotional style.
In this work, we address zero-shot, any-to-any speaker identity conversion with unseen source and target speakers, using only a short reference utterance at inference time in a non-parallel setting without aligned utterances between source and target speakers.

Data-driven VC systems traditionally rely on parallel speech data, where pairs of utterances share the same linguistic content but differ in speaker identity. Early approaches \cite{chen2014voice, sun2015voice} learned direct mappings from such aligned data. However, collecting large-scale parallel corpora is expensive and impractical, motivating the development of non-parallel VC methods.
A typical strategy is to disentangle content and speaker information from source and target speech, respectively, and use them to reconstruct converted speech.

A widely used approach extracts speaker-independent linguistic units from ASR models, such as Phonetic PosteriorGrams (PPGs) \cite{sun2016phonetic} and phonemes \cite{liu2021any}, or features from TTS models \cite{zhang2021transfer, park2020cotatron}, and then synthesizes speech using speaker-dependent features of the target speaker. While these methods achieve strong content preservation, they are language-dependent and require either manually annotated text data or highly accurate ASR/TTS systems.

Other works attempt to disentangle content and speaker information without linguistic supervision \cite{qian2019autovc, chou2019ada-in}. To extend to any-to-any VC, where both source and target speakers are unseen during training, subsequent methods leverage self-supervised learning (SSL) models \cite{polyak2021speech, lin2021fragmentvc, huang2022s3prl}. To further reduce speaker-related attributes, various strategies are applied, including bottleneck constraints \cite{li2023freevc}, vector quantization \cite{wu2020vqvc+}, and instance normalization \cite{chen2021again}. However, disentanglement remains imperfect, and learned representations may leak speaker information or fail to preserve sufficient content, leading to degraded conversion quality.

Recent work has proposed alternatives to disentanglement-based approaches by synthesizing data for one or both sides of the source--target pair. This is achieved either by using TTS models to generate aligned source and target segments from different speakers \cite{tu2025o_ovc}, or by applying existing VC models to convert audio from the target to the source domain \cite{liu2024seedvc, guo2025synthvc}. 

Another alternative to disentanglement-based methods is to use a simple k-nearest neighbor (KNN) strategy to map between SSL features \cite{baas2023knnvc}. Frame-level SSL representations are extracted from both source and target speech, and each source frame is replaced by its most similar target frame, followed by waveform reconstruction using a vocoder. While simple and effective, this approach suffers in one-shot scenarios, where limited target data leads to sparse neighbors and degraded intelligibility and quality. The Phoneme Hallucinator \cite{shan2024phoneme} attempts to mitigate this issue by generating synthetic target representations to expand the neighbor pool.

In this work, we propose a voice conversion framework that leverages controlled synthetic data generation during training. Specifically, we synthesize training inputs using a KNN-based retrieval scheme that preserves linguistic content while altering speaker identity, enabling scalable supervision from non-parallel speech. Our system consists of a pretrained SSL WavLM encoder \cite{chen2022wavlm}, a transformer-based latent converter, and a HiFi-GAN vocoder \cite{kong2020hifi}, trained using a combination of reconstruction, adversarial, and speaker identity losses. Importantly, we incorporate a speaker verification model \cite{desplanques2020ecapa} as waveform-level supervision, which directly enforces speaker similarity and leads to consistent improvements over prior methods.

To summarize, the main contributions of this work are:

\begin{itemize}
    \item We propose a non-parallel, zero-shot, any-to-any voice conversion framework that leverages controlled synthetic data generation for scalable supervision without aligned speech data.
    
    \item We introduce a speaker verification--based waveform-level loss that directly optimizes speaker similarity, leading to consistent improvements in identity preservation.
    
    \item We demonstrate superior performance in speaker similarity and equal error rate compared to recent state-of-the-art systems, while remaining comparable in WER, CER, and subjective quality evaluations.
    
    \item We show strong multilingual generalization, achieving effective cross-lingual performance without any fine-tuning on non-English data.
\end{itemize}

\begin{table*}[!t]
\centering
\caption{Performance on English data. Our method outperforms other methods in Speaker Similarity and Equal Error Rate (EER), while remaining comparable in WER/CER and subjective evaluation. Best results from each prompt duration catetory is highlighted in bold.}
\footnotesize            
\setlength{\tabcolsep}{2.5pt}
\renewcommand{\arraystretch}{0.9}

\begin{tabular*}{\textwidth}{@{\extracolsep{\fill}}lccccccc}
\toprule
\multirow{2}{*}{Model} & \multicolumn{5}{c}{Objective} & \multicolumn{2}{c}{Subjective} \\
\cmidrule(lr){2-6}\cmidrule(lr){7-8}
 & Prompt Dur. (s) & Spk Sim ($\uparrow$) & EER ($\uparrow$) & WER ($\downarrow$) & CER ($\downarrow$) & MOS ($\uparrow$) & SMOS ($\uparrow$) \\
\midrule

\multirow{4}{*}{Seed-VC \cite{liu2024seedvc}} 
& 3  & 0.455 & 0.007 & 0.041 & 0.019 & 3.545 & 3.200 \\
& 10 & 0.578 & 0.060 & 0.042 & 0.021 & 3.000 & 3.500 \\
& 30 & 0.622 & 0.083 & 0.026 & 0.011 & 3.666 & 3.583 \\
& 60 & 0.630 & 0.090 & 0.033 & 0.014 & 3.400 & \textbf{3.600} \\

\midrule
\multirow{4}{*}{KNN-VC \cite{baas2023knnvc}}  
& 3  & 0.380 & 0.033 & 0.430 & 0.277 & 1.182 & 1.500 \\
& 10 & 0.552 & 0.053 & 0.115 & 0.062 & 2.588 & 2.642 \\
& 30 & 0.617 & 0.070 & 0.042 & 0.018 & 3.500 & 3.500 \\
& 60 & 0.631 & 0.073 & 0.046 & 0.020 & 3.000 & 2.800 \\

\midrule
\multirow{4}{*}{Vevo \cite{zhang2025vevo}}    
& 3  & 0.510 & 0.047 & \bf{0.038} & \bf{0.013} & \bf{3.909} & \bf{4.100} \\
& 10 & 0.639 & 0.133 & \bf{0.033} & \bf{0.011} & \bf{3.765} & \bf{4.071} \\
& 30 & 0.651 & 0.093 & \bf{0.022} & \bf{0.007} & \bf{3.750} & \bf{3.750} \\
& 60 & 0.332 & 0.010 & 0.197 & 0.129 & 1.300 & 1.200 \\

\midrule
\multirow{4}{*}{O\_O-VC \cite{tu2025o_ovc}} 
& 3  & 0.368 & 0.010 & 0.050 & 0.018 & 3.636 & 3.500 \\
& 10 & 0.419 & 0.020 & 0.036 & 0.019 & 3.588 & \bf{4.071} \\
& 30 & 0.431 & 0.007 & 0.038 & 0.013 & 3.583 & 3.666 \\
& 60 & 0.453 & 0.010 & \bf{0.031} & \bf{0.012} & \bf{3.800} & 3.100 \\

\midrule
\multirow{4}{*}{Ours}    
& 3  & \bf{0.612} & \bf{0.097} & 0.056 & 0.023 & 3.455 & 3.300 \\
& 10 & \bf{0.713} & \bf{0.180} & 0.049 & 0.022 & 3.647 & 4.000 \\
& 30 & \bf{0.717} & \bf{0.170} & 0.047 & 0.019 & 3.666 & 3.583 \\
& 60 & \bf{0.712} & \bf{0.150} & 0.050 & 0.021 & \textbf{3.800} & 3.300 \\

\end{tabular*}
\label{tab:quantitative_results}
\end{table*}

\section{Method}

An overview of our method is shown in \Cref{fig:overview}. We introduce an end-to-end training framework for any-to-any zero-shot voice conversion under a non-parallel data setting.
The core motivation of this design is to enable supervised learning of speaker conversion without requiring parallel or aligned speech, by introducing a synthetic intermediate representation.
Given a corpus of speech segments labeled by speaker identity, we generate training triplets composed of a reference segment $A_2$, a synthetic source segment $\hat{B}_1$, and a real target segment $a_1$. During training, the model learns to map the synthetic source to the acoustic characteristics of the reference speaker, while using the real target as supervision.
At inference time, however, the system operates directly on real (non-synthetic) input speech, and the model generalizes from synthetic training inputs to real-world conversion.
The framework relies as reference solely on the input segment, and therefore naturally supports any-to-any zero-shot voice conversion.

\subsection{Palindromic Training Framework}

The proposed framework enables training from non-parallel speech data by constructing synthetic source–target pairs in a palindromic manner. Instead of relying on aligned utterances across speakers, the framework generates synthetic source features from real target speech and trains the model to invert this mapping back to the target. As a result, the system enables effective voice conversion in a zero-shot setting and requires only a short reference segment from the target speaker at inference time, without requiring parallel corpora.

The backbone of our training scheme is inherited from the work of KNN-VC \cite{baas2023knnvc}, which presents a straightforward yet effective approach for any-to-any voice conversion. In this method, features of a source speaker are matched to those of a target speaker via simple k-nearest neighbors (KNN) retrieval, leveraging the strong representations of a pre-trained self-supervised feature extractor \cite{chen2022wavlm}. For optimal results, however, this approach requires a relatively large corpus of reference features: while reasonable quality can be achieved with approximately five minutes of target speech, naturalness and artifact reduction continue to improve as more data is provided (e.g., ten minutes or more).

Given a segment of a target speaker $a_1$ and a gallery of an arbitrary reference speaker $b_{ref}$, we extract features $A_1$, $B_{ref}$ and generate a synthetic feature set using KNN-VC, denoted by $\hat{B}_1$. This process can be performed offline.

Using the pair $(\hat{B}_1, a_1)$, we treat $\hat{B}_1$ as a synthetic source and $a_1$ as the real supervised target. Together with an additional reference feature set from the same target speaker, denoted by $A_2$, this triplet provides a palindromic training framework for the transformer, which is trained to convert the synthetic source features into target-speaker features, producing $\widetilde{B}_1$. The details of the training process are presented in the next section.

\subsection{Training Procedure}

Training is performed in three stages. First, we train a vocoder to map WavLM features to audio waveforms. Next, we train the transformer-based conversion model. Finally, we train a vocoder on the converted features produced by the transformer.

\noindent\textbf{Stage 1: Vocoder Pre-training.}\quad
In the first stage, we train a vocoder to reconstruct audio waveforms from WavLM \cite{chen2022wavlm} features in an auto-encoding manner. The vocoder is optimized using a combination of waveform-based reconstruction losses, including the Multi-Resolution STFT (MR-STFT) loss \cite{yamamoto2020parallel}, together with adversarial training using multi-period (MPD) and multi-scale (MSD) discriminators \cite{kong2020hifi}. This stage is crucial for enabling stable optimization of the transformer in the following stage, where supervision is applied at the waveform level.

\noindent\textbf{Stage 2: Transformer Training.}\quad
In the second stage, we train the transformer model to perform feature-level voice conversion under the proposed palindromic training framework. The training objective consists of two components: an L1 loss between the predicted features and the ground-truth target features, and a speaker loss at the waveform level that enforces speaker identity consistency.\\
\indent The speaker loss is computed using a pre-trained speaker verification model \cite{desplanques2020ecapa}. Specifically, we extract speaker embeddings and hidden representations from both the reference waveform $a_2$ and the converted waveform $\widetilde{a}_1$, and compute the cosine similarity and L1 distance between them. This loss plays a central role in the conversion process, as it directly encourages the converted output to match the target speaker's identity. We find it crucial to use the same reference segment as input to both the transformer and the speaker loss in order to obtain stable and consistent training.

\noindent\textbf{Stage 3: Vocoder Post-training.}\quad
In the final stage, we train a vocoder on the converted features produced by the transformer, following the same palindromic training framework as in the previous stage. The vocoder is optimized using the same objectives as in the first stage. This stage allows the vocoder to adapt specifically to the feature distribution produced by the transformer, resulting in improved naturalness and reduced artifacts in the final synthesized speech.

\section{Experiments}



In Stage~1, following KNN-VC~\cite{baas2023knnvc}, we extract features from the 6th layer of WavLM~\cite{chen2022wavlm} and train a 16M-parameter vocoder for 100K steps. 
In Stage~2, we train a 77M-parameter six-layer Transformer with 16 attention heads and a hidden dimension of 1024 for 800K steps using Adam~\cite{Kingma2015adam} with a learning rate of $3 \times 10^{-4}$. 
In Stage~3, a new instance of the vocoder is trained for 100K steps.

\newsavebox{\sharedlegendbox}
\definecolor{skyblue}{RGB}{86,180,233}

\pgfplotsset{
  compat=1.18,
  durationModern/.style={
    ybar,
    bar width=4.4pt,
    enlarge x limits=0.22,
    width=\linewidth,
    height=0.9\linewidth,
    axis line style={draw=none},
    tick style={draw=none},
    tick label style={font=\scriptsize},
    label style={font=\scriptsize},
    ymajorgrids,
    grid style={line width=0.25pt, draw=black!12},
  },
    knnStyle/.style ={fill=orange!90!black, draw=none},
    oovcStyle/.style={fill=skyblue!80!black, draw=none},
    vevoStyle/.style={fill=magenta!75!black, draw=none},
    oursStyle/.style={fill=blue!85!black,    draw=none}, 
    seedStyle/.style={fill=green!70!black,   draw=none},
}

\begin{lrbox}{\sharedlegendbox}
\begin{tikzpicture}
\begin{axis}[
  hide axis,
  legend to name=fig:shared-legend,
  legend columns=5,
  legend style={
    draw=none,
    /tikz/every even column/.append style={column sep=6pt},
    font=\scriptsize
  },
  legend cell align=left,
  legend image code/.code={
    \draw[#1, draw=none] (0cm,-0.1cm) rectangle (0.28cm,0.18cm);
  },
]
\addplot+[oovcStyle] coordinates {(0,1)}; \addlegendentry{OOVC}
\addplot+[knnStyle]  coordinates {(0,1)}; \addlegendentry{KNN-VC}
\addplot+[vevoStyle] coordinates {(0,1)}; \addlegendentry{Vevo}
\addplot+[oursStyle] coordinates {(0,1)}; \addlegendentry{Ours}
\addplot+[seedStyle] coordinates {(0,1)}; \addlegendentry{Seed-VC}
\end{axis}
\end{tikzpicture}
\end{lrbox}

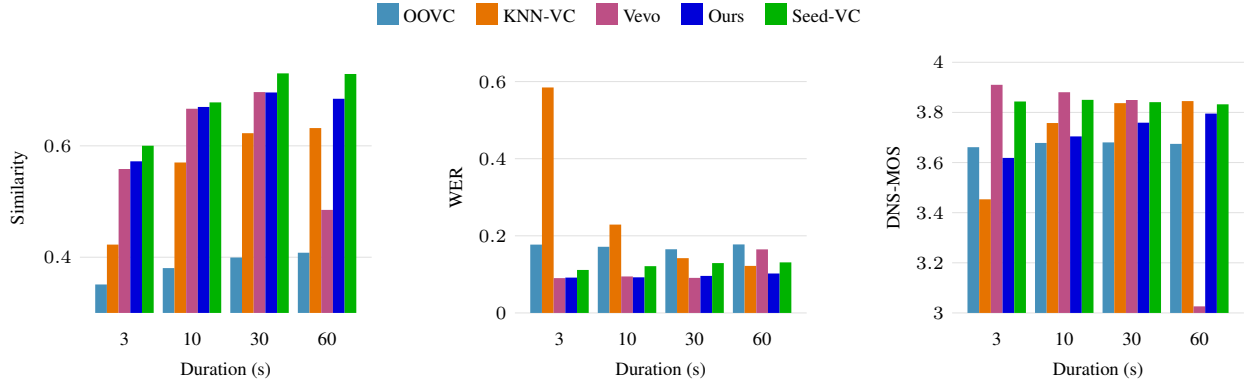
\begin{figure*}[t]
\centering

\pgfplotslegendfromname{fig:shared-legend}
\vspace{0.4em}

\begin{subfigure}[t]{0.32\textwidth}
\centering
\begin{tikzpicture}
\begin{axis}[
  durationModern,
  ylabel={Similarity},
  xlabel={Duration (s)},
  symbolic x coords={3,10,30,60}, xtick=data,
  ymin=0.3, ymax=0.75,
]
\addplot+[oovcStyle, bar shift=-2*\pgfplotbarwidth] coordinates {(3,0.3511) (10,0.3805) (30,0.3995) (60,0.4081)};
\addplot+[knnStyle,  bar shift=-1*\pgfplotbarwidth] coordinates {(3,0.4227) (10,0.5701) (30,0.6227) (60,0.6319)};
\addplot+[vevoStyle, bar shift= 0*\pgfplotbarwidth] coordinates {(3,0.5584) (10,0.6666) (30,0.6965) (60,0.4850)};
\addplot+[oursStyle, bar shift= 1*\pgfplotbarwidth] coordinates {(3,0.5722) (10,0.6698) (30,0.6958) (60,0.6845)};
\addplot+[seedStyle, bar shift= 2*\pgfplotbarwidth] coordinates {(3,0.6002) (10,0.6780) (30,0.7301) (60,0.7290)};
\end{axis}
\end{tikzpicture}
\end{subfigure}
\hfill
\begin{subfigure}[t]{0.32\textwidth}
\centering
\begin{tikzpicture}
\begin{axis}[
  durationModern,
  ylabel={WER},
  xlabel={Duration (s)},
  symbolic x coords={3,10,30,60}, xtick=data,
  ymin=0.0, ymax=0.65,
]
\addplot+[oovcStyle, bar shift=-2*\pgfplotbarwidth] coordinates {(3,0.1772172036) (10,0.1715547180) (30,0.1651286454) (60,0.1776774004)};
\addplot+[knnStyle,  bar shift=-1*\pgfplotbarwidth] coordinates {(3,0.5847072048) (10,0.2291202236) (30,0.1420730643) (60,0.1219779187)};
\addplot+[vevoStyle, bar shift= 0*\pgfplotbarwidth] coordinates {(3,0.09029453461) (10,0.09460679424) (30,0.09094831773) (60,0.1648299950)};
\addplot+[oursStyle, bar shift= 1*\pgfplotbarwidth] coordinates {(3,0.09165810439) (10,0.09237349495) (30,0.09602830027) (60,0.1022786170)};
\addplot+[seedStyle, bar shift= 2*\pgfplotbarwidth] coordinates {(3,0.1114852624) (10,0.1212793117) (30,0.1292902128) (60,0.1311261654)};
\end{axis}
\end{tikzpicture}
\end{subfigure}
\hfill
\begin{subfigure}[t]{0.32\textwidth}
\centering
\begin{tikzpicture}
\begin{axis}[
  durationModern,
  ylabel={DNS-MOS},
  xlabel={Duration (s)},
  symbolic x coords={3,10,30,60}, xtick=data,
  ymin=3.0, ymax=4.0,
]
\addplot+[oovcStyle, bar shift=-2*\pgfplotbarwidth] coordinates {(3,3.661229054) (10,3.678175597) (30,3.680367958) (60,3.674477123)};
\addplot+[knnStyle,  bar shift=-1*\pgfplotbarwidth] coordinates {(3,3.453307095) (10,3.757645403) (30,3.836977675) (60,3.84514577)};
\addplot+[vevoStyle, bar shift= 0*\pgfplotbarwidth] coordinates {(3,3.9101117) (10,3.880524374) (30,3.849525202) (60,3.026215088)};
\addplot+[oursStyle, bar shift= 1*\pgfplotbarwidth] coordinates {(3,3.618482397) (10,3.70427403) (30,3.75896598) (60,3.7953351)};
\addplot+[seedStyle, bar shift= 2*\pgfplotbarwidth] coordinates {(3,3.843546549) (10,3.850114981) (30,3.840892111) (60,3.832304966)};
\end{axis}
\end{tikzpicture}
\end{subfigure}

\caption{Effect of prompt duration across multiple languages. Speaker Similarity ($\uparrow$) is shown on the left, WER ($\downarrow$) in the center, and DNS-MOS ($\uparrow$) on the right. Our method achieves the best WER while remaining comparable to other approaches in terms of speaker similarity and DNS-MOS.}
\label{fig:multilingual-perf}
\end{figure*}

\subsection{Datasets}

We use the LibriSpeech dataset \cite{panayotov2015librispeech} for training and testing in English. For multi-lingual testing we use test sets in the Multilingual LibriSpeech dataset \cite{Pratap2020MLSAL}. Training was done by randomly pairing audio segments of different source and target speakers, generating three different target speakers for each segment, across 960 hours of the LibriSpeech. For precise evaluation, in each language we evaluate performance across four different duration of the prompt signal, across three seeds of fifty randomly selected source-target audio pairs. The exact number of speakers and total duration of each language's evaluation set is depicted in \Cref{tab:per-lang-n-speakers}.

\begin{table}[h]
\centering
\footnotesize
\caption{Speakers per language in evaluation sets.}
\setlength{\tabcolsep}{4pt}
\begin{tabular}{lccc}
\toprule
Lang. & \#Source & \#Target & Total Duration (min.) \\
\midrule
Dutch      & 6  & 6  & 148 \\
English    & 27 & 27 & 85  \\
French     & 17 & 17 & 150 \\
German     & 25 & 25 & 151 \\
Italian    & 10 & 10 & 153 \\
Polish     & 4  & 4  & 147 \\
Portuguese & 10 & 10 & 156 \\
Spanish    & 18 & 18 & 152 \\
\bottomrule
\end{tabular}
\label{tab:per-lang-n-speakers}
\end{table}

\subsection{Evaluation Metrics}

\noindent\textbf{Objective Evaluation.}\quad
We evaluate performance using Speaker Similarity, Equal Error Rate (EER), Word Error Rate (WER), and Character Error Rate (CER), and assess multi-language quality using DNS-MOS \cite{reddy2022dnsmos}.


We use Redimnet Speaker Verifier \cite{yakovlev24_redimnet} to compute Speaker similarity and EER.
Note, that for a fair comparison we use a speaker verification module different than the one we use for optimizing our model \cite{desplanques2020ecapa}.
We use Whisper-Large-V3 \cite{radford2022whisper} to transcribe converted segments for WER and CER evaluation.

\noindent\textbf{Subjective Evaluation.}\quad
For human evaluation, we rate naturalness using the Mean Opinion Score (MOS) and rate speaker similarity using the Speaker Similarity Mean Opinion Score (SMOS). Both metrics rate on a 1-5 scale.
Three samples were randomly selected from each duration category, resulting in a total of twelve samples. Each category was evaluated by at least ten participants.

\subsection{Results and Analysis}

We compare our method against four recent voice conversion systems \cite{baas2023knnvc, liu2024seedvc, zhang2025vevo, tu2025o_ovc}, including KNN-VC, which serves as the backbone of our approach. This allows us to assess both absolute performance and the contribution of our modifications over a strong baseline.

Quantitative results under the English setting are presented in \Cref{tab:quantitative_results}. Our method consistently outperforms all competing systems in speaker similarity and EER across all prompt duration settings, demonstrating improved speaker identity preservation and verification robustness. In terms of intelligibility and perceptual quality metrics (WER, CER, MOS, and SMOS) our performance remains comparable to state-of-the-art systems, indicating that the gains in speaker similarity do not come at the expense of naturalness or transcription accuracy.

Performance across non-English languages is shown in \Cref{fig:multilingual-perf}. Notably, no fine-tuning was performed on non-English data. Despite this, our approach achieves consistent improvements in WER across all prompt duration settings, indicating strong cross-lingual generalization. Furthermore, speaker similarity and DNS-MOS remain comparable to competing systems, suggesting that our method maintains speaker consistency and perceptual quality even in unseen linguistic conditions.

We provide detailed analysis of our approach and audio samples across all languages in the project page\footnote{\url{https://palindromic-vc.github.io/}}.

Compared to the backbone of our method, KNN-VC \cite{baas2023knnvc}, our approach demonstrates consistently stronger performance across both objective and subjective metrics, for both English and non-English data. The performance gap is particularly pronounced in the low-resource setting with only 3 seconds of target speaker reference, where KNN-VC degrades substantially while our method remains robust and significantly outperforms it across all evaluation measures.

This trend is further confirmed by the subjective results. Under short prompt conditions (3 and 10 seconds), our model achieves significantly higher MOS and SMOS scores than KNN-VC, indicating superior perceptual quality and speaker similarity when limited reference information is available. As the prompt duration increases (30 and 60 seconds), the performance gap narrows, but our method continues to maintain a consistent advantage, demonstrating stronger stability and scalability with respect to reference length.

Vevo \cite{zhang2025vevo} and Seed-VC \cite{liu2024seedvc} are trained on substantially larger datasets (60K and 100K hours, respectively), whereas our model uses roughly 3K hours. The Seed-VC \cite{liu2024seedvc} examples on our demo page indicate challenges in preserving source prosody. OOVC \cite{tu2025o_ovc} addresses prosodic mismatch with a dedicated F0 conditioning adapter. In contrast, our approach requires no explicit prosody modeling, yet effectively maintains prosodic consistency between source and converted speech.

\subsection{Ablation}



                         






\begin{table}[!t]
\centering
\caption{Ablation study of vocoder post-training, demonstrating improved DNS-MOS with preserved speaker similarity and WER.}

\footnotesize
\resizebox{\columnwidth}{!}{%
\begin{tabular}{lcccccc}
\toprule
 Model & Dur. & Spk Sim ($\uparrow$) & WER ($\downarrow$) & DNS-MOS ($\uparrow$) \\
\midrule

\multirow{4}{*}{\shortstack{Without Vocoder\\Post-training}}   & 3   & 0.580 & 0.057 & 3.396 \\
                                  & 10  & 0.696 & 0.046 & 3.621 \\
                                  & 30  & 0.718 & 0.037 & 3.753 \\
                                  & 60  & 0.725 & 0.033 & 3.801 \\
\midrule

\multirow{4}{*}{\shortstack{With Vocoder\\Post-training}}   & 3   & 0.582 & 0.050 & 3.366 \\
                                  & 10  & 0.689 & 0.042 & 3.770 \\
                                  & 30  & 0.695 & 0.033 & 3.819 \\
                                  & 60  & 0.697 & 0.045 & 3.834 \\
                         
\end{tabular}
}
\label{tab:ablation}
\end{table}



                         

We conduct an ablation study to evaluate the impact of vocoder post-training, and report the results in \Cref{tab:ablation}. The model was trained specifically for this controlled ablation setting. Audio samples are provided in the project page\footnotemark[1].
Before the post-training stage, the converted audio exhibits noticeable auditory artifacts, reflected in the low DNS-MOS scores, despite achieving high speaker similarity and low WER. Incorporating vocoder post-training effectively mitigates these artifacts, as evidenced by consistently higher DNS-MOS scores for three out of four durations, while maintaining comparable speaker similarity and WER.

\section{Conclusions \& Future Work}

We present a non-parallel, zero-shot, any-to-any voice conversion framework that enables supervised speaker conversion without aligned speech. By introducing a palindromic training strategy based on controlled KNN-generated synthetic features, our method leverages non-parallel data to learn speaker identity mapping in a scalable, data-efficient manner.

A key contribution is the incorporation of a waveform-level speaker verification loss, which directly enforces speaker similarity and leads to consistent improvements in speaker similarity and EER, without degrading intelligibility or perceptual quality.

Experiments on LibriSpeech and Multilingual LibriSpeech demonstrate that our approach outperforms recent state-of-the-art systems in speaker identity preservation while remaining comparable in WER, CER, MOS, and DNS-MOS. Notably, strong cross-lingual generalization is achieved without any non-English fine-tuning.

These results show that controlled synthetic supervision combined with explicit speaker-level objectives provides an effective alternative to traditional disentanglement-based voice conversion methods.

For future work, we aim to adapt the proposed method to large-scale datasets and highly expressive speech, as well as real-time and streaming voice conversion.

\clearpage

\section{Generative AI Use Disclosure}
A generative AI tool was used to assist with language editing and refinement of specific sections of this manuscript. It was not used to generate substantial technical content, research ideas, experimental design, or results.

\bibliographystyle{IEEEtran}
\bibliography{mybib}

\end{document}